# PREDICTING HETEROGENEOUS TREATMENT EFFECTS OF BUILDING ENERGY SAVING RETROFITS USING CAUSAL MACHINE LEARNING

*Completed Research Paper*

Kevin Zalipski, Paderborn University, Paderborn, Germany, kevin.zalipski@upb.de

David Zapata Gonzalez, Paderborn University, Paderborn, Germany, david.zapata@upb.de

Oliver Müller, Paderborn University, Paderborn, Germany, oliver.mueller@upb.de

## Abstract

*Information Systems research increasingly relies on machine learning (ML) to predict outcomes in complex sociotechnical systems, yet predictive models are not designed to identify causal effects. This limitation is particularly critical in building retrofits, where unbiased estimates of energy savings are essential for climate policy and investment decisions. Because retrofit adoption is shaped by household and building characteristics that also affect energy consumption, predictive ML can yield biased effect estimates. This paper systematically benchmarks leading causal ML estimators, including metalearners (S-, T- and X-Learners) and DoubleML across multiple retrofit interventions. To enable this comparison, we construct a physically grounded simulation in which true treatment effects and realistic adoption biases are known. Results show that DoubleML achieves the lowest estimation errors, particularly for complex envelope retrofits. These findings demonstrate that orthogonalising the treatment assignment improves causal effect estimation and provides a methodological foundation for large-scale energy retrofit and policy evaluation.*

## 1 Introduction

Information Systems (IS) research has rapidly adopted machine learning (ML) methods to model socio-technical phenomena using large-scale, high-dimensional data, primarily to improve prediction performance (Abdel-Karim et al., 2021). Yet many questions central to the IS discipline are inherently causal, for example: the effect of introducing a new system feature, changing a platform policy or deploying an IT-enabled intervention (Athey, 2017). Standard predictive ML methods cannot answer such questions because they optimise for correlational accuracy rather than causal identification. This gap has motivated a rapidly expanding IS literature on causal ML, which combines the flexibility of ML with principled causal inference to estimate treatment effects and support causal reasoning in complex socio-technical settings (Kuzmanovic et al., 2024; Tafti & Shmueli, 2020, 2025; von Zahn et al., 2025; Zapata Gonzalez et al., 2025). This limitation of standard predictive ML is particularly critical in energy informatics contexts such as building energy retrofits, where treatment assignments (e.g., which house receives the treatment) are highly biased.

Buildings account for a substantial share of global energy use and carbon emissions. According to the International Energy Agency (IEA) (2023), their operations represent around 30% of global final energy consumption and 26% of energy-related $CO_2$ emissions. In most countries, the residential sector is a major contributor to this impact (Eurostat, 2025; U.S. Energy Information Administration (EIA), 2024). Energy retrofitting, through measures such as façade insulation, window replacement or heating-system upgrades, is therefore central to national and global decarbonisation strategies (Camarasa et al., 2022). Accurate estimation of the energy savings achievable through such retrofits is critical, since these estimates form the foundation for cost-benefit analysis, risk assessment and the prioritization of retrofit options.

Traditionally, retrofit evaluations rely on physical building-energy simulations, which model thermal processes in detail but require extensive data, expert knowledge and computational resources (Ali et al., 2021; Jensen et al., 2022; Ma et al., 2012). Moreover, simulation outcomes often diverge from actual realised savings because occupant characteristics, local weather variation and building heterogeneity are difficult to capture precisely. Data-driven approaches have recently emerged as scalable alternatives: by learning patterns from historical retrofit data, they can predict post-retrofit energy use with less modelling effort. However, such predictive models typically optimise for correlation-based prediction accuracy and therefore fail to reveal causal effects (von Zahn et al., 2026), that is, what the energy use would have been had the building not been renovated.

Causal Machine Learning (ML) addresses this limitation by estimating the causal impact of interventions while controlling for confounding variables (Kaddour et al., 2022; Pearl, 2014). This allows for a root-cause analysis of observed effects, counterfactual analysis and answering of what-if questions (Blöbaum et al., 2024; Sharma & Kiciman, 2020), as demonstrated by recent IS applications in domains such as development aid allocation, e-commerce and modelling of dynamical systems (Kuzmanovic et al., 2024; von Zahn et al., 2025; Zapata Gonzalez et al., 2025).

To advance this line of research, we develop a structurally grounded simulation of building energy retrofits in which the true heterogeneous treatment effects and empirically motivated adoption biases are explicitly modelled and therefore known. This design enables direct benchmarking of causal estimators against a verified ground truth and allows for a systematic evaluation of their robustness under realistic, non-random treatment assignment conditions that closely mirror observed retrofit adoption patterns. By doing so, we generate quantitative evidence on how accurately causal ML methods recover heterogeneous treatment effects in the presence of confounding. While simulation-based evaluation necessarily depends on the specified data-generating process, it uniquely permits the controlled assessment of estimator performance under known structural relationships and bias mechanisms, an assessment that is fundamentally infeasible with purely observational data.

Accordingly, our study is guided by the following research question:

**RQ:** How accurately can causal ML methods recover heterogeneous treatment effects of building retrofits when treatment assignment is biased?

To answer the research question, we evaluate different causal ML methods by comparing their estimated heterogeneous treatment effects to the ground-truth effects. This research makes three primary contributions. First, we provide the first quantitative assessment of causal ML estimator accuracy for retrofit scenarios in individual houses with known bias, thereby complementing existing empirical applications that can only assess aggregate plausibility at the average treatment effect (ATE) level. Second, we systematically benchmark OLS, metalearners and DoubleML, and identify their relative strengths and weaknesses in recovering treatment effects. Third, we demonstrate that advanced causal ML methods, especially DoubleML, yield more credible effect estimates for energy-efficiency interventions, strengthening the empirical basis for data-driven retrofit policy design.

# 2 Background

## 2.1 Building retrofitting and energy saving models

Household retrofitting comprises a broad range of interventions that reduce a building's energy demand and improve thermal comfort. Measures include active systems, such as ventilation upgrades and modern heating or cooling installations, and passive envelope improvements, such as thermal insulation, airtightness enhancement and window replacement (Recart & Dossick, 2022; Saffari &

Beagon, 2022). Typically, retrofit projects follow a five-step process. First, the project's scope, targets and resources are defined, supported by a pre-retrofit survey. Next, an energy audit and building performance assessment identify areas for improvement. This is followed by the evaluation and prioritization of retrofit options based on energy-saving potential, economic feasibility and associated risks. The fourth phase involves on-site implementation and commissioning of selected measures to ensure optimal building performance. Finally, post-measurement and verification validate energy savings, complemented by occupant feedback to assess satisfaction (Ma et al., 2012).

Energy-saving assessments are commonly performed through simulation-based tools. These start from a calibrated baseline model of the building, apply retrofit measures virtually and estimate post-retrofit demand before verification and final recalibration (Dermentzis et al., 2019). Regarding the software, approaches generally fall into two categories. The first category uses simple thermodynamical formulas (Kirschbaum et al., 2025) or simplified algorithms such as the Passive House Planning Package (PHPP) (Passive House Institute, 2015), which is based on Microsoft Excel. This tool is straightforward to use and can provide reasonable results when assessing retrofit measures (Dermentzis et al., 2019; Moran et al., 2014). The second category includes studies that employ detailed dynamic simulation tools such as EnergyPlus (National Renewable Energy Laboratory (NREL), 2025) and its graphical front-ends (Beagon et al., 2020; Chuah et al., 2013; Crawley et al., 2001; Garg et al., 2020). These provide finer physical detail and modelling of buildings, potentially resulting in more accurate representations of building performance, but also require a higher level of expertise, greater effort and thus significantly more time setting up each individual building model.

In response, ML approaches have gained traction for retrofit modelling (Ascione et al., 2017; Deb et al., 2021; Nguyen et al., 2024; Thrampoulidis et al., 2021). ML algorithms learn from historical input-output data to forecast outcomes for future unseen instances from the same distribution (James et al., 2013). These models are highly effective for prediction tasks and have the advantage of a simpler modelling process compared to physical simulations. However, ML models primarily rely on simple correlations between variables (Bontempi & Flauder, 2015; Pearl, 2009) and often fail to capture the underlying causal structure of the data-generating process. As a result, they are vulnerable to overfitting and confounding bias, particularly in cases where variables simultaneously influence both the treatment (e.g., retrofit measures) and the outcome (e.g., energy consumption). For example, houses located in colder regions may be both more likely to undergo retrofitting and to exhibit higher energy consumption, even without any retrofit taking place. If location-specific factors such as the climate severity are not adequately controlled for, the model may falsely attribute these differences to the retrofit itself. Such hidden confounders distort estimated causal effects and compromise the models' ability to reliably assess the true impact of specific retrofit interventions, often resulting in an over- or underestimation of the actual effect in real world scenarios.

## 2.2 Causal ML

These limitations of correlation-based ML motivate the use of causal inference. One foundational approach is the potential outcome framework (POF), also known as the Neyman-Rubin causal model (Rubin, 1974), which provides a rigorous basis for estimating causal effects. A causal effect for a unit is defined as the difference between its two potential outcomes: what would happen if it received a specific treatment ($Y(1)$) versus what happens if it did not ($Y(0)$). Because only one outcome can be observed, it is known as the "fundamental problem of causal inference" (Holland, 1986).

In practice, the causal effects can be estimated based on the average treatment effect (ATE) by using the average of the outcomes of the treated and untreated groups. Experimental designs, such as randomized controlled trials (RCTs), ensure that the treatment assignment is not systematically related to other factors that might influence the outcome, thereby eliminating confounding variables. Similarly, quasi-experimental designs attempt to mimic randomization in settings where actual random assignment is not feasible, often by leveraging natural experiments (Cook et al., 2002).

As experimental designs are impractical or infeasible in many real-world scenarios, including our case of building retrofitting (i.e., it is economically and ethically not feasible to randomly assign retrofit treatments to existing buildings), practitioners must often rely on observational data. However, even from observational data, causal effects can be estimated under certain assumptions, such as when the relevant confounders influencing both the treatment and the outcome are measured (assumption of unconfoundedness or ignorability), when the treated units have similar probabilities of receiving the treatment (overlap or common support assumption) and when one unit's outcome is not affected by another unit's treatment (SUTVA) (Imbens & Rubin, 2015; Lee, 2016). In such situations, one option is to use statistical techniques such as matching to compare treated units with non-treated counterparts that exhibit similar covariates but differ only in the treatment variable, enabling estimation of the ATE as the difference in outcomes between groups (Cunningham, 2021; Imbens & Rubin, 2015). Another option is to use propensity score matching (PSM) to estimate the probability of the treatment based on observed covariates and form balanced groups for comparison. Additionally, regression-based adjustment methods can be used to control for confounders. In practice, multiple approaches are often combined to strengthen causal inference.

Importantly, treatment effects are rarely uniform across units. They vary with individual or contextual characteristics, giving rise to heterogeneous treatment effects (HTEs) or Conditional Average Treatment Effects (CATEs). These values quantify the expected effect of a treatment $T$ on an outcome $Y$ conditional on a set of observed individual characteristics or features, denoted as $X$ (e.g., retrofit effect on houses with a specific wall thickness and construction year). Theoretically, simple statistical models can be used to calculate the CATE, such as a linear regression with control variables. However, this often falls short in capturing the heterogeneity of treatment effects due to its inherent assumptions of linearity and its struggle to model complex non-linear relationships and interactions without extensive manual feature engineering (Molak, 2023). To overcome this, we can integrate flexible ML algorithms into methods of causal inference for estimating causal effects, allowing them to automatically capture the complex ways treatment effects vary across different individuals (Molak, 2023). This integration enables explicit deconfounding and provides quasi-counterfactual capabilities for predicting individualized responses, which are crucial for decision-making at the building level, such as estimating the impact of a specific retrofit on a specific house.

## 2.3 Related work

Early applications of causal machine learning in the IS literature demonstrate its value for rigorously estimating causal effects. For instance, Kuzmanovic et al. (2024) propose a causal ML framework to estimate HTEs of continuous development aid and use it to identify cost-effective allocation strategies that could substantially reduce global HIV infection rates. Von Zahn et al. (2025) use causal ML and digital footprints to personalize green nudges in e-commerce. By targeting customers with information on the environmental impact of returns, they reduced returns and increased profits while avoiding backfire effects. Similarly, Zapata Gonzalez et al. (2025) combine causal discovery with human domain expertise to model data centre operations for prediction and "what-if" analysis, emphasizing how causal ML can bridge data-driven and theory-driven modelling.

Recent research in the energy domain has begun applying these causal approaches to large-scale building datasets. For example, Jiang & Kazmi (2025) investigate data-driven approaches for modelling building thermal dynamics for control-oriented applications and demonstrate the limitations of using traditional ML approaches, and how causal ML models perform better in out-of-distribution scenarios and in attributing causal effects. In addition, X. Chen et al. (2022) propose a framework for asking "what-if" questions in the initial building design processes that highlight how causal discovery and causal graphs significantly help to identify variables that impact the heating load of a building. They found that models trained with these causally relevant features show an improved performance compared to other data-driven methods.

Other studies have also applied causal methods specifically to energy-retrofit data. Weber et al. (2025) implement the potential outcome framework on a national building-stock dataset containing more than 100,000 building-year observations of Swiss houses. They employ metalearners (S-, T-, X- and R-learners) with various base models, including neural networks, gradient boosting and linear regression, to estimate both ATEs and CATEs for different retrofit measures. Their findings demonstrate the practical feasibility of causal ML in this domain and show that neural-network-based learners yield the most stable results. However, because their data is observational, true counterfactual outcomes are unobservable, restricting evaluation to the ATEs rather than measurable accuracy against known CATEs for the individual houses. D'Amico et al. (2025) pursue a complementary causal approach, using directed acyclic graphs (DAGs) to analyse the causal structure of retrofit effects across 22.6 million households in England. While their framework effectively identifies plausible causal pathways and supports both ATE and CATE estimation, the validation relies on falsifiability tests rather than quantitative verification of effect accuracy, again due to the lack of counterfactual ground truth.

# 3 Methodology

## 3.1 Dataset creation

To generate a comprehensive dataset that allows us to train and evaluate the CATE predictions, we adapt thermodynamic formulas from Kirschbaum et al. (2025). The simulation design reflects realistic variance in environmental and dwelling characteristics, while allowing controlled variation of the retrofit treatment parameters.

The following equations are used:

- transmission heat loss of the building: $\phi_{TL,build} = \sum_k A_k * U_k * f_{x,k} * (\theta_{int,build} - \theta_e)$
- air heat loss: $\phi_{AL,build} = V_{build} * n_{build} * \rho_{air} * c_{p,air} * (\theta_{int,build} - \theta_e)$
- resulting heat load: $\phi_{HL,build} = \phi_{TL,build} + \phi_{AL,build}$
- annual heating requirement: $Q_{h,build} = \phi_{HL,build} * t_h$

For building element k, $A_k$ is the component area, $U_k$ the heat transfer coefficient (U-value) and $f_{x,k}$ the temperature correction factor. The building volume is defined as: $V_{build} = living_area * 2.5$. The building air change rate $n_{build}$ is calculated as: $0.4h^{-1}$ (which is the user ventilation rate) plus an infiltration rate that depends on the building decade. $\rho_{air} * c_{p,air}$ describes the product of the air density and the specific thermal capacity of the air. $(\theta_{int,build} - \theta_e)$ indicates the difference in outside and inside norm temperatures, while $t_h$ is the variable for the yearly load hours. U-values for windows, roofs, floors and exterior walls are drawn from the exemplary single-family home specification of Loga et al. (2015). The "current state" U-values are used for the pre-retrofit consumption and the "modernisation package 2: future-oriented U-values" for the post-retrofit consumption.

| Class | Period | Wall U-value | Roof U-value | Window U-value | Floor U-value | User air change rate in $h^{-1}$ | Infiltration rate in $h^{-1}$ |
|---|---|---|---|---|---|---|---|
| A | 1850-1859 | 2,0 / 0,14 | 2,6 / 0,14 | 2,8 / 0,8 | 2,9 / 0,27 | 0,4 | 1 |
| B | 1860-1918 | 1,7 / 0,13 | 1,3 / 0,14 | 2,8 / 0,8 | 1,2 / 0,23 | 0,4 | 0,9 |
| C | 1919-1948 | 1,7 / 0,13 | 1,4 / 0,14 | 2,8 / 0,8 | 1,0 / 0,23 | 0,4 | 0,8 |
| D | 1949-1957 | 1,4 / 0,13 | 1,4 / 0,14 | 2,8 / 0,8 | 1,0 / 0,23 | 0,4 | 0,7 |
| E | 1958-1968 | 1,2 / 0,13 | 0,8 / 0,14 | 2,8 / 0,8 | 1,6 / 0,25 | 0,4 | 0,6 |
| F | 1969-1978 | 1,0 / 0,13 | 0,5 / 0,09 | 2,8 / 0,8 | 1,0 / 0,23 | 0,4 | 0,5 |
| G | 1979-1983 | 0,8 / 0,12 | 0,5 / 0,14 | 4,3 / 0,8 | 0,8 / 0,21 | 0,4 | 0,4 |
| H | 1984-1994 | 0,5 / 0,11 | 0,4 / 0,14 | 3,2 / 0,8 | 0,6 / 0,20 | 0,4 | 0,2 |

*Table 1. House characteristics per class; for the U-values, the first value represents the pre-retrofitting status, while the second value indicates the post-retrofitting value.*

Table 1 summarises the parameters used per building period class. For the geometry (living area, window share, ratio of floor, roof and façade), U-values, user ventilation factors, difference of inside and outside norm temperatures, and full load hours of each house, a small random variation is applied to obtain as diverse data as possible. The initial setup regarding the base variables can be seen in the Appendix. The randomness factors used for the different variables and house classes can be seen in a GitHub repository[1], containing the Python-Notebook with the complete code used for the data creation process.

The considered treatments are:

- windows only, where only the windows are improved
- windows and roof, where both the windows and the roof are renovated
- exterior walls only, where just the exterior walls are retrofitted
- windows, roof and exterior walls, where all these entities are improved at once

To generate the simulation data allowing us to get the actual treatment effect (true CATE[2]), only the U-values are altered via the parameter mapping introduced above, depending on the chosen treatment. All other input variables regarding the location, house and residential characteristics are held constant between pre- and post-retrofit simulations. This allows us to generate perfectly paired counterfactuals that differ only in thermal envelope performance and therefore isolate the causal consumption reduction attributable to the retrofit treatment.

## 3.2 Treatment bias

Treatment assignment is deliberately biased in our simulation to reflect the empirically observed non-random adoption of energy-efficiency retrofits. We construct the bias parameters from the central regularities reported in recent observational studies. The literature jointly highlights (i) socio-economic determinants (income, household characteristics), (ii) building stock characteristics (age, size) and (iii) climatic conditions.

Across studies (Belaïd & Massié, 2023; Eakins et al., 2025; Schleich et al., 2021; Walsh, 1989), higher income or higher standards of living generally increase the probability of retrofit adoption, particularly for envelope measures such as wall insulation or double-glazed window installation, though there are exceptions for specific measures. For example, retrofits are income-insensitive in the study by Filippini & Kumar (2024) and Ossokina et al. (2021) document that monetary framings of retrofit information can decrease retrofit acceptance in tenants, while comfort framing has the opposite effect. Several studies also show building age gradients: older buildings tend to be more likely to be renovated (Filippini & Kumar, 2024; Peñasco & Anadón, 2023), with retrofit likelihood peaking for mid-century houses (ca. 1940-1970) and increasing with age (Schleich et al., 2021). On the other hand, one study showed that houses built between 1975 and 1999 are significantly more likely to get wall, roof and window improvements than houses built before 1949 (Belaïd & Massié, 2023). Building size tends to depress adoption probability (Filippini & Kumar, 2024), but other studies imply a higher adoption likelihood with increased building size (Eakins et al., 2025; Walsh, 1989). Climatic signals are mixed. In some studies, colder locations with more heating degree days and fewer cooling degree days increase conservation-oriented improvements (Walsh, 1989), while in others, there is no effect, apart from specifically influencing roof insulations in regions with a slightly lower temperature compared to Mediterranean or very cold regions (Belaïd & Massié, 2023).

These findings are implemented as structured multivariate probability shifts in our treatment assignment mechanism. Accordingly, retrofit treatments in our synthetic observational dataset are not independent of income, location or building age. Crucially, these variables not only influence the probability of receiving a retrofit but also affect energy consumption directly, for example, through climate conditions or age-related characteristics of the house, like the U-values. The resulting dependencies create genuine confounding in the data creation process, not just treatment selection bias. In Figure 1, a simplified DAG based on the data-creation process illustrates this structure: multiple determinants simultaneously affect both the treatment and the outcome, mirroring the mechanisms documented in empirical studies. By embedding these confounding pathways explicitly into the simulation, the dataset provides a realistic and challenging testbed against which the treatment effect estimations can be evaluated.

---

[1] Link to the repository: https://github.com/Zalipski/ecis_causal_retrofits

[2] Strictly speaking, the "true CATE" in our setting corresponds to an individual treatment effect (ITE), since both potential outcomes are known for each dwelling. We retain the CATE terminology for consistency with the causal ML literature and to align the ground-truth quantity with the estimators evaluated in this study.

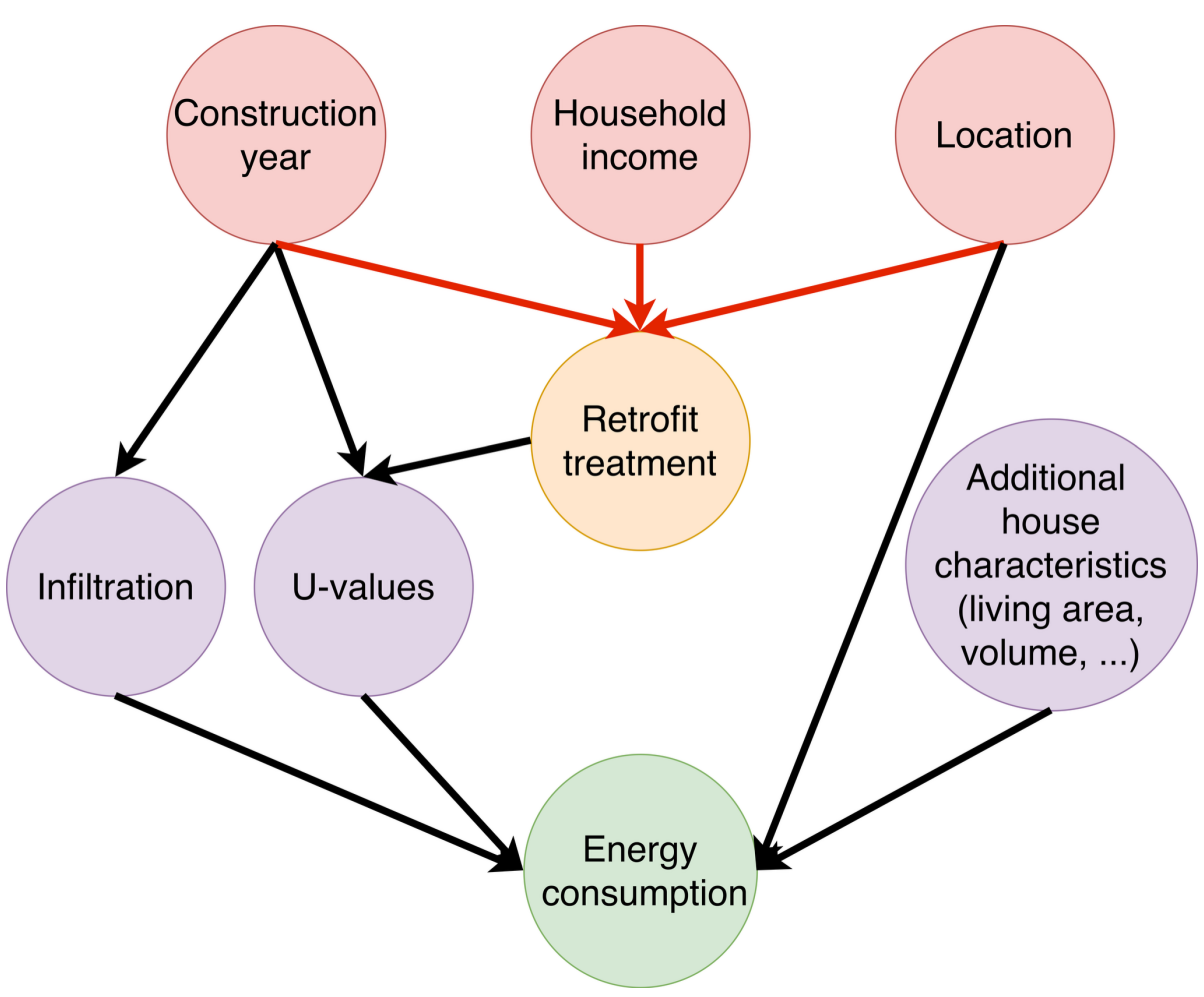

*Figure 1. DAG showing the confounding present in the data.*

| **Treatment** | **# Houses** | **Rel. # Houses** | **Effect in kWh** | **Location (% in cold climate)** | **Income in USD** | **Construction year** |
|---|---|---|---|---|---|---|
| No retrofit | 11,865 | 0.237 | 0.0 (0.0) | 44.2 | 40,880 (10,887) | 1977.4 (14.8) |
| Windows only | 6,256 | 0.125 | -2,020.4 (1,306.7) | 50.9 | 53,448 (6,904) | 1966.5 (19.4) |
| Windows + roof | 9,300 | 0.186 | -5,975.5 (4,196.9) | 49.2 | 55,270 (16,316) | 1950.6 (38.4) |
| Exterior walls only | 7,884 | 0.158 | -12,696.0 (8,276.5) | 51.6 | 54,543 (8,289) | 1914.1 (47.0) |
| Windows + roof + exterior walls | 14,695 | 0.294 | -21,500.3 (13,154.1) | 53.6 | 50,163 (18,455) | 1913.7 (43.0) |

*Table 2. Statistics for the different treatments; the columns regarding the effect, location, income and construction year show the mean and the standard deviation in parentheses.*

As visible in Table 2, more thorough retrofits (exterior wall only, and windows, roof and exterior wall) tend to occur in older buildings, in colder locations and in households with higher average incomes, i.e., the expected directions implied by the literature synthesis above. Thus, the dataset spans a realistic range of treatment heterogeneity and treatment selection bias, approximating what actual observational energy-performance data would be like.

The distribution of CATE values per treatment in the training data can be seen in the mentioned GitHub repository. All treatments exhibit substantial heterogeneity in energy savings. Most buildings experience a relatively moderate reduction in energy consumption, while a very small number benefits from very large savings.

## 3.3 Data assumptions

Before applying causal ML methods, the necessary assumptions must be addressed to ensure valid treatment effect estimates: the stable unit treatment value assumption (SUTVA), the unconfoundedness and the overlap, or common support, assumption. The stable unit treatment value assumption (SUTVA) requires that each unit's outcome depends only on its own treatment assignment and that there are no hidden variations of the treatment (Rubin, 1980). In our setting, SUTVA holds by design, as each building's energy consumption is determined solely by its own characteristics and retrofit status, with no interference between buildings, and each treatment is well-defined through explicit U-value mappings. The unconfoundedness assumption (that all relevant covariates are included in the set of variables used for modelling) is not directly testable; it must be argued on theoretical or empirical grounds (Imbens & Rubin, 2015). Given our experimental design and known data-generation process, the treatment assignment is by construction independent of the potential outcomes conditional on the observed covariates. Therefore, we can assume unconfoundedness holds in this setting, meaning that after conditioning on the covariates, there are no unobserved confounders jointly affecting the treatment assignment and the outcome.

The common support assumption requires that each unit has a non-zero probability of receiving every treatment, given its covariates (Imbens & Rubin, 2015). This can be examined visually by assessing the degree of overlap in treatment probabilities. Since the model includes multiple covariates, we compute each unit's propensity score, the estimated probability of receiving a given treatment conditional on observed characteristics (Caliendo & Kopeinig, 2008).

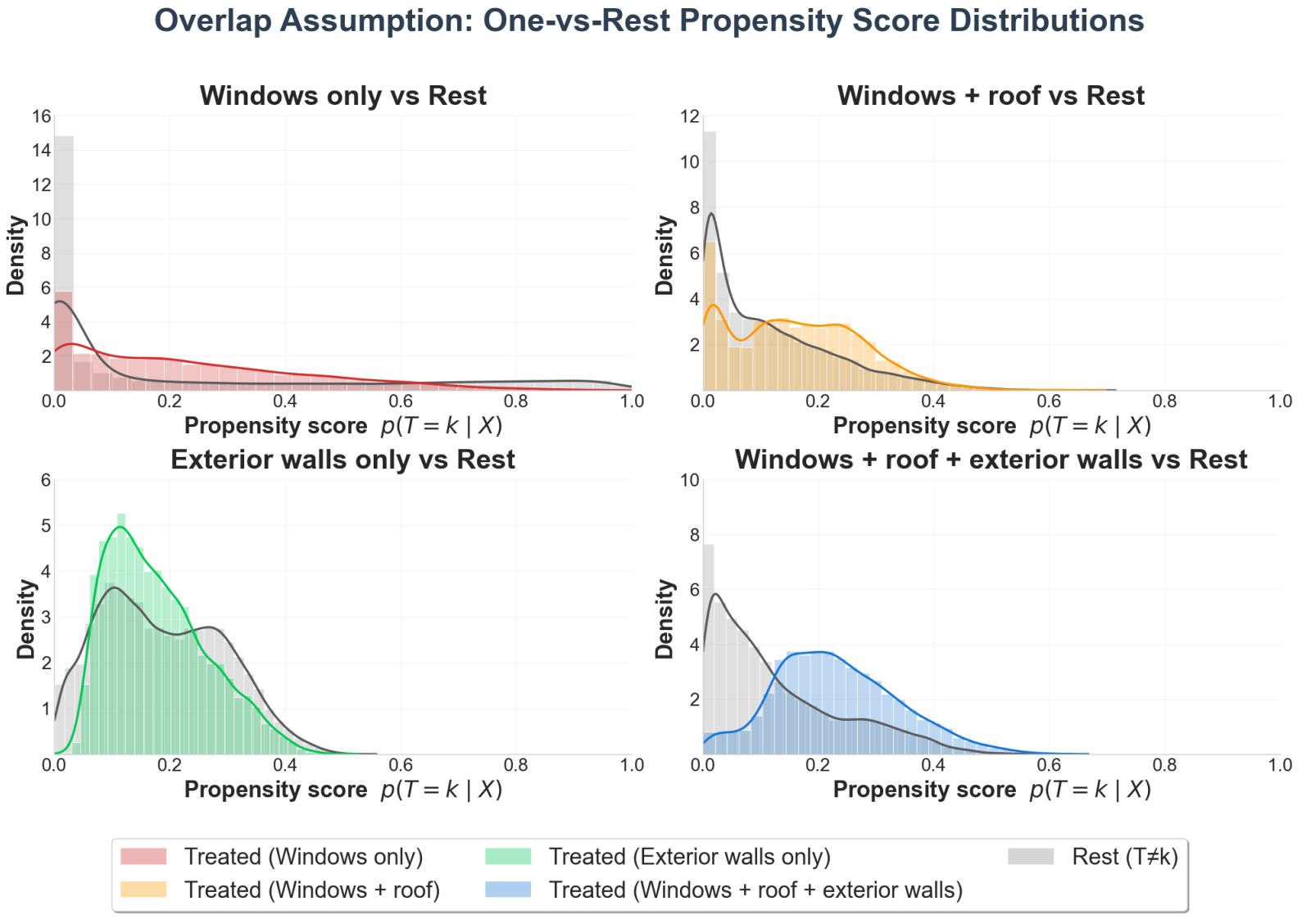

*Figure 2. Overlap assumption per treatment.*

Figure 2 displays the one-vs-rest propensity score distributions (histograms with density overlays) for each treatment. Each plot compares the treated group with the corresponding control group (gray). Across all four panels, the distributions show substantial overlap across a broad range of propensity scores, with no treatment appearing nearly deterministic (i.e., scores concentrated solely near 1 for treated units or near 0 for the rest) (Imbens & Rubin, 2015). This overlap indicates adequate common support, providing evidence that the common support assumption holds for all treatments.

## 3.4 Models

All models are trained in a binary setup, meaning that for each treatment, only the corresponding treatment group is used for training and testing; the control group is the same across all treatment-model combinations. The feature set includes pre-retrofit energy consumption, household size, income, construction year, location, pre-retrofit U-values of the floor/windows/roof/exterior walls and full load hours. Due to the household size being highly correlated with house size, floor/windows/roof/exterior wall area and building volume, and because pre-retrofit energy consumption is strongly correlated with both transmission and air heat load, as well as a strong correlation of the location with the difference between norm outside and inside temperature being present, only the first-mentioned variable in each of these correlated groups is retained.

**OLS**: A linear regression (OLS) is included as a naive reference point to quantify the bias that arises when treatment effects are estimated without correcting for confounding. It estimates the average treatment effect (ATE) for each treatment, representing a common first step in applied research. It is implemented via the Python statsmodels package (Seabold & Perktold, 2010).

**Metalearners**: The metalearners used are the S-, T- and X-Learner. Metalearners represent a class of causal ML techniques that are model-agnostic and flexible in estimating CATEs. They can be paired with any ML algorithm to evaluate treatment effects in a principled way (Curth & Van der Schaar, 2021; Künzel et al., 2019). The simplest metalearner is the S-Learner (single learner). The S-Learner uses a single base learner (a traditional ML model) to model the relationship between the features, treatment and outcome. The CATE is then estimated by calculating the difference between a prediction for a unit with the treatment and another without the treatment (Künzel et al., 2019).

However, an S-Learner uses the treatment like any other feature and when treatment effects are small relative to other features, the model may ignore it. A T-Learner (two learner) addresses this limitation by estimating separate response functions for the treated and untreated groups. The data is first split into two subsets based on the treatment indicator: one containing all observations with the treatment being 1 and one with the treatment set to 0. A separate model is then trained on each subset to learn the outcome function for that group. At prediction time, the CATE is obtained by subtracting the predicted control outcome from the predicted treated outcome for the same feature vector (Künzel et al., 2019). This approach ensures that treatment-specific outcome patterns are captured, even when the effect of the treatment is small.

Building on this concept, the X-Learner goes a step further by imputing counterfactuals for each group and then modelling the individual treatment effects separately. It combines these estimates using a weighting function, often derived from the propensity score, to balance confidence across treatment groups. This makes the X-Learner especially effective in scenarios with imbalanced treatment assignment (Künzel et al., 2019).

**Double Machine Learning**: Another important causal ML method is Double Machine Learning (DoubleML) (Chernozhukov et al., 2018), also referred to as debiased machine learning. The core idea relies on the Frisch-Waugh-Lovell (FWL) theorem (Frisch & Waugh, 1933) and is utilised to debias the treatment effect estimations. It involves training an outcome model (predicting the outcome from control variables) and a treatment model (predicting the treatment from the same control variables). The residuals from both models represent the variation in the outcome and the treatment that can't be explained by the confounders alone. Regressing the outcome model residuals on the treatment model residuals yields an estimate of the causal effect that is orthogonal to the confounding structure, thereby removing regularisation bias introduced by flexible ML models. This approach, combined with cross-fitting (repeatedly splitting the data), provides theoretical guarantees for debiasing causal effect estimations while allowing for the use of arbitrary ML models as base learners.

## 3.5 Hyperparameter Tuning

All model-treatment combinations are tuned using RandomSearchCV of the sklearn library (Pedregosa et al., 2011) for 30 iterations with the search space displayed in Table 3 and a cv value of 3. The search space is selected primarily based on the default values in the sklearn library for the random forest model class. The final parameters of the models can be seen in Table 4. The previously mentioned GitHub repository also contains the implementation code for the training and evaluation of the models. The X-Learner and the DoubleML models are implemented via the Python EconML library (Battocchi et al., 2019) and all models use a Random Forest as the base learner, implemented via the Python sklearn library (Pedregosa et al., 2011).

| Parameter | Values |
|---|---|
| # Estimators | 50, 100, 150 |
| Max. depth | 5, 10, - |
| Min. samples leaf | 5, 10, 15 |
| Max. features | -, sqrt, log2 |

*Table 3. Search space for hyperparameter tuning; "-" indicates a None value.*

| Model | # Estimators | Max. depth | Min. samples leaf | Max. features |
|---|---|---|---|---|
| S-Learner | 100, 100, 100, 100 | -, -, -, - | 5, 5, 5, 5 | -, -, -, - |
| T-Learner | 100, 50, 100, 100 | -, -, -, - | 5, 5, 5, 5 | -, -, -, - |
| X-Learner | 100, 100, 100, 100 | 10, -, -, 10 | 5, 5, 5, 10 | -, -, -, - |
| DoubleML Regressor | 100, 100, 100, 100 | 10, -, -, 10 | 5, 5, 5, 10 | -, -, -, - |
| DoubleML Classifier | 150, 100, 100, 50 | 5, 10, -, 5 | 10, 10, 10, 5 | -, log2, sqrt, - |

*Table 4. Final model parameters; the four values per cell correspond to the four treatments in order; "-" indicates a None value.*

# 4 Results

Table 5 summarises the mean absolute error (MAE) and normalized MAE (MAE divided by the average true CATE value) of the CATE estimates produced by the different estimators across the different treatments (i.e., retrofit types). OLS yields by far the highest overall error (5,701.51; nMAE: 62.0%), confirming the need for both an explicit causal adjustment and heterogenous effect estimation.

DoubleML yields the lowest average error across all treatments (800.62; nMAE: 11.6%), followed by the T-Learner (872.96; nMAE: 11.4%) and the X-Learner (949.87; nMAE: 11.5%), while the S-Learner performs substantially worse (1,134.43; nMAE: 13.7%).

| Treatment | OLS (ATE) | S-Learner | T-Learner | X-Learner | DoubleML |
|---|---|---|---|---|---|
| Windows only | 1,717.43 (83.4%) | 429.36 (20.8%) | 395.58 (19.2%) | **376.60 (18.3%)** | 471.52 (22.9%) |
| Windows + Roof | 3,495.08 (58.7%) | 718.64 (12.1%) | 639.64 (10.7%) | 610.44 (10.2%) | **564.11 (9.5%)** |
| Exterior Walls only | 7,265.98 (57.2%) | 1,831.19 (14.4%) | 1,278.69 (10.1%) | 1,324.83 (10.4%) | **1,159.04 (9.1%)** |
| Windows + Roof + Exterior Walls | 10,327.52 (48.8%) | 1,558.53 (7.4%) | 1,177.94 (5.6%) | 1,487.63 (7.0%) | **1,007.81 (4.8%)** |
| Average | 5,701.51 (62.0%) | 1,134.43 (13.7%) | 872.96 (11.4%) | 949.87 (11.5%) | **800.62 (11.6%)** |

*Table 5. MAE for the different model and treatment combinations; best model based on the MAE in bold; normalized MAE in parentheses.*

However, the ranking is not uniform across individual treatments. For the smallest retrofit (windows only), the X-Learner is most accurate (376.60; nMAE: 18.3%), followed closely by the T-Learner with an MAE of 395.58 (nMAE: 19.2%). DoubleML achieves the worst performance of all CATE models (471.52; nMAE: 22.9%), 94.92 above the X-Learner. For windows and roof, DoubleML achieves the lowest error (564.11; nMAE: 9.5%), slightly outperforming the X- and T-Learners (610.44 (nMAE: 10.2%) and 639.64 (nMAE: 10.7%), respectively) by 46.33 and 75.53. As retrofit complexity and true effect magnitude increase, the performance gap between DoubleML and the other estimators widens. In the exterior wall only treatment, DoubleML's MAE (1,159.04; nMAE: 9.1%) is 119.65 lower than the T-Learner (1,278.69; nMAE: 10.1%) and 165.79 lower than the X-Learner (1,324.83; nMAE: 10.4%). This advantage becomes more pronounced for the comprehensive windows, roof and exterior walls case, where DoubleML (1007.81; nMAE: 4.8%) undercuts the T-Learner (1,177.94; nMAE: 5.6%) by 170.13 and the X-Learner (1,487.63; nMAE: 7.0%) by 479.82.

The scatterplots in Figure 3 visually corroborate these error patterns. Across all treatments, the S- and T-Learners deviate substantially from the diagonal once treatment effects exceed the moderate range observed in the windows-only case. Across the more extensive retrofit types, the metalearners consistently attenuate large effects and never predict CATE values below roughly -50.000 to -60.000 kWh, even though the true CATE distribution for the training data contains values well beyond this range. As mentioned earlier, relatively large savings per treatment occur only in a small fraction of training samples and the metalearners appear unable to extrapolate into this sparsely populated region, leading to the flattening observed in the scatterplots. The X-Learner shows a similar but less pronounced attenuation, particularly for the exterior wall only and the comprehensive retrofit.

In contrast, DoubleML tracks the diagonal line closely across all treatments and effect sizes, with only minor deviations for windows, roof and exterior wall only. For this comprehensive retrofit case, its predictions remain almost perfectly aligned with ground truth, deviating only slightly when the treatment effects become very large. This indicates that DoubleML is able to generalise reliably even from the limited number of observations in the high-saving regime, further underscoring its robustness in recovering large CATE values under biased treatment assignment.

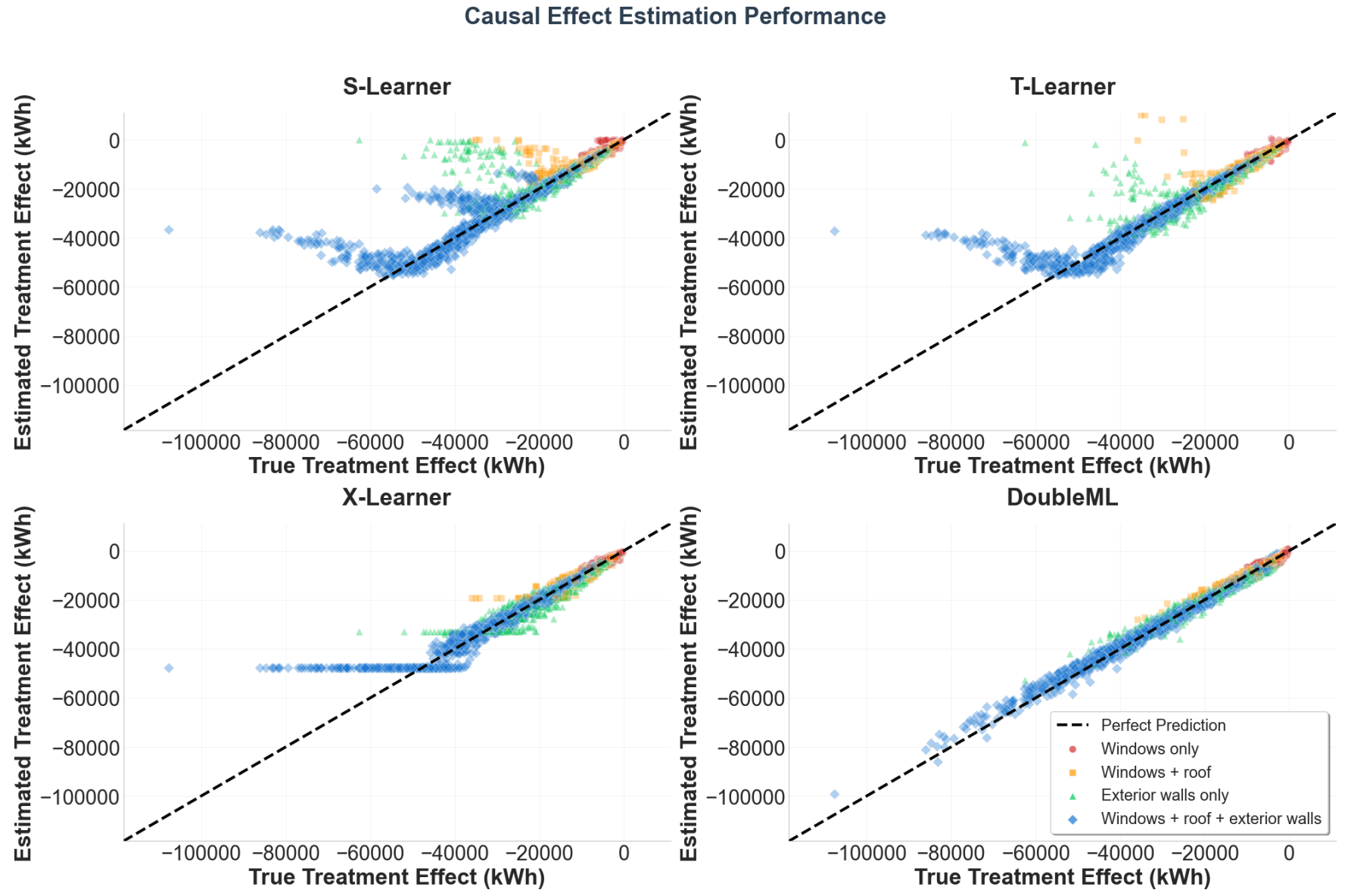

*Figure 3. Estimated vs actual treatment effect.*

A sample size analysis (Table 6) shows that model performance improves substantially as sample sizes increase from 1,000 to 10,000 total observations, with diminishing returns beyond 30,000. At the smallest sample size, all models exhibit high errors and DoubleML in particular requires more data to outperform the metalearners. From approximately 10,000 observations onward (corresponding to roughly 1,200-3,000 per treatment), DoubleML achieves the lowest MAE, and its advantage stabilises at 30,000 observations.

| Samples per treatment | OLS | S-Learner | T-Learner | X-Learner | DoubleML |
|---|---|---|---|---|---|
| 120-266 (1,000 total) | 5856.26 | 5075.70 | 3217.02 | **2033.82** | 4480.21 |
| 603-1,509 (5,000 total) | 5649.19 | 2039.15 | **1268.10** | 1287.61 | 1516.87 |
| 1,239-2,970 (10,000 total) | 5704.19 | 1735.43 | 1366.40 | 1203.24 | **943.52** |
| 3,725-8,905 (30,000 total) | 5703.02 | 1318.18 | 1048.58 | 980.19 | **756.74** |

*Table 6. Average MAE across all treatments for different sample sizes; best model in bold.*

A SHAP analysis of the individual model predictions reveals that confounding variables, especially construction year, remain influential in the metalearners' CATE estimates, whereas DoubleML effectively removes their influence through orthogonalisation. Figure 4 illustrates this for the comprehensive treatment.

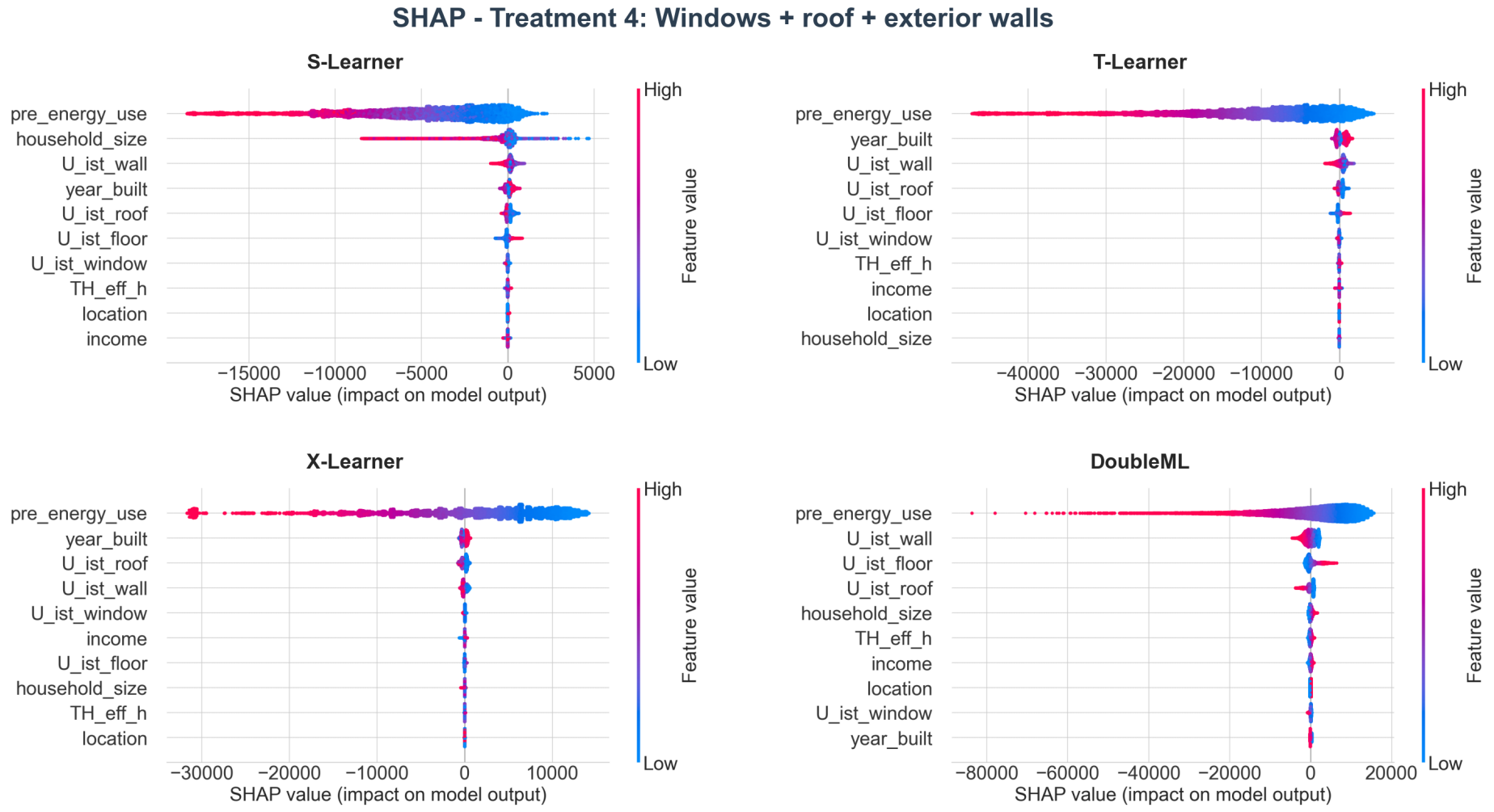

*Figure 4. SHAP beeswarm plots for the comprehensive treatment across all models.*

# 5 Discussion

The results demonstrate that causal ML estimators are not only theoretically sound but also effective within our simulation-based evaluation for estimating retrofit-related energy savings under realistic treatment bias. The large gap between OLS and the more advanced causal estimators quantifies the bias introduced by ignoring selective treatment assignments. These findings underline that, in observational settings where treatment adoption is selective, models that explicitly control for confounding produce considerably more reliable estimates than those simply regressing outcomes on treatment indicators, particularly for deeper and more complex retrofit packages where confounding and effect heterogeneity are strongest. The results also show that causal ML methods, particularly DoubleML, are a suitable tool for both researchers and practitioners to obtain retrofit effect estimations. The methods allow saving a considerable amount of time compared to using complex simulations like EnergyPlus, while still getting accurate estimations within our experimental setting.

Beyond overall accuracy, the comparative behaviour of the estimators offers insight into how model architecture interacts with retrofit complexity. For the smallest intervention (windows only), the T-Learner achieves the lowest MAE, followed closely by the X-Learner and DoubleML. This outcome is consistent with theoretical expectations: when treatment effects are moderate and the untreated population is large, metalearners that train separate models for treated and untreated units can approximate counterfactual outcomes efficiently. As the retrofit complexity increases, however, DoubleML becomes the most stable and accurate approach across all remaining interventions. For the deeper envelope retrofits, such as exterior wall only and windows, roof and exterior walls, DoubleML's ability to orthogonalise the treatment assignment clearly improves estimation accuracy. Moreover, DoubleML generalises more reliably into the high-saving tail despite the scarcity of observations in that region. A key structural reason is that metalearners rely entirely on tree-based base learners, which cannot extrapolate beyond the range of their training data, as predictions are bounded by the leaf-node averages of the observed sample. DoubleML, by contrast, uses a linear regression on the residuals at its final stage, which is not subject to this boundary constraint and can therefore produce effect estimates beyond the range seen during training. Its robustness in these high-effect, high-confounding scenarios is particularly relevant for practical policy contexts, as such retrofits require credible effect estimation for efficient resource allocation.

The present results align with recent empirical work demonstrating the value of causal ML for evaluating building retrofits. Weber et al. (2025) and D'Amico et al. (2025) have demonstrated that causal learning frameworks can be deployed successfully at scale, revealing heterogeneity in retrofit performance across large building populations. Our simulation-based approach complements this empirical stream by providing an environment in which the true causal effects are known. This makes it possible to quantify not only which estimators perform best overall, but also under what conditions, such as varying levels of confounding or treatment intensity, different causal ML approaches begin to diverge. The ability to measure these performance gradients is essential for building trust in causal inference methods before deploying them to large-scale policy evaluation.

Causal ML relies on data assumptions and violations of these can bias the results. In real-world applications, unconfoundedness must be argued. Although capturing all possible confounders is rarely feasible, we recommend practitioners to use comprehensive and relevant variables proposed in this study, such as housing characteristics, energy use, infiltration rates, retrofit treatments, construction year, household income (or a proxy thereof) and location, can make this assumption reasonable (Imbens & Rubin, 2015). For the common support assumption, practitioners can examine propensity score distributions for treated and control units empirically (Caliendo & Kopeinig, 2008) and trim observations in regions with poor overlap to improve validity. SUTVA is likely to hold in retrofit settings, as a building's energy consumption is generally unaffected by whether neighbouring buildings are retrofitted and treatments are typically well-defined through specific material and component specifications.

The importance of comprehensive covariate measurement is further illustrated by our SHAP analysis, which shows that metalearners do not fully eliminate the influence of confounding variables on their CATE estimates. DoubleML's orthogonalization more effectively removes this confounding influence, but the advantage depends on relevant confounders being included in the model. Our sample size analysis illustrates that model performance improves substantially up to approximately 1,200-3,000 observations per treatment, beyond which returns diminish. Notably, DoubleML requires more data than the metalearners to realise its advantage, but consistently achieves the lowest errors from that threshold onward.

As shown in the results, causal effect estimation errors are expected when applying causal ML models. Our simulation is intentionally complex, including randomness, non-linear relationships, interactions and biases that make estimation challenging. Since SUTVA, overlap and unconfoundedness hold by design, remaining discrepancies mainly reflect this complexity. For practitioners who use real observational data, it is important to consider that errors may also result from violations of these assumptions, the complexity of the problem, the quality of collected features and limited data, particularly interventional data.

Our results directly contribute to the emerging IS literature on causal ML by demonstrating the importance of explicitly modelling the treatment assignment process to recover unbiased heterogeneous treatment effects (Kuzmanovic et al., 2024; von Zahn et al., 2025). We show that ignoring a selective treatment assignment systematically biases estimated intervention effects in complex socio-technical systems. Moreover, our empirical evaluation provides concrete evidence that causal estimators, particularly DoubleML, outperform correlation-based models under strong selection bias and out-of-distribution scenarios.

The results suggest three broader lessons from a methodological standpoint for modelling complex socio-technical systems for IS scholars and practitioners. First, the accuracy of causal estimators depends as much on explicitly modelling the treatment assignment process as on the richness of the available variables. When the decision is inherently selective, in our case, the retrofit decision, which is influenced by factors such as household income, location or building age that simultaneously influence the energy consumption, ignoring these mechanisms can systematically bias estimated savings and lead to misleading conclusions about policy effectiveness. Second, estimator choice should be guided by the nature of the intervention and data at hand. While metalearners may provide competitive results for simple, low-signal treatments, DoubleML, which explicitly orthogonalises treatment and outcome models, offers superior robustness when confounding is strong or when policy decisions depend on strong treatments with large heterogeneity effects. Third, to actually draw the best performance, practitioners should try to use as much relevant variables as available and possible.

Naturally, this study is limited in several aspects. It does not capture the full spectrum of biases that manifest in real-world observational energy performance data. For example, we do not model behaviour-induced rebound/comfort-taking effects or behavioural conservation responses, despite evidence that such elasticities exist and often differ by socio-economic strata (Eakins et al., 2025; Umit et al., 2019). Second, additional selection confounders beyond the biases we integrate (income, location, building age) also drive adoption: household-level heterogeneity in education, debt aversion, access to capital, energy price exposure, age, household composition and tax incentive regimes all materially shift retrofit likelihoods (Belaïd & Massié, 2023; Eakins et al., 2025; Filippini & Kumar, 2024; Schleich et al., 2021). Third, even after adoption, occupant interaction with the building and its HVAC system (e.g., window opening, thermostat settings, presence schedules) substantially affects energy consumption outcomes (S. Chen et al., 2021). In our simulation, this is either held static or omitted. While we intentionally impose empirically grounded treatment assignment bias, real-world settings are plausibly even more confounded. Accordingly, our evaluation remains a conservative approximation of the observational complexity encountered in practice. Finally, we only utilise Random Forest as the base learner for the causal ML models and do not include other causal models like the R- and DR-Learner or Causal Forests.

Future research could extend the bias mechanisms and behavioural heterogeneity embedded in the simulation, such as incorporating rebound effects, capital constraints or occupant decision dynamics. Furthermore, combining the benchmark approach with real-world observational datasets could calibrate simulation realism and strengthen external validity. Additionally, since both data generation and model training involve stochasticity, formal significance testing across multiple independent dataset replications would further strengthen the conclusions and represents a natural extension for future work. Researchers could also explore alternative base learners, such as XGBoost or neural networks, and extend the estimator comparison to include other causal models.

# 6 Acknowledgements

This study was supported by the Federal Ministry of Economics and Climate Protection (BMWK) of Germany under Grant No. 03EN6024A-G.

# References

Ali, U., Shamsi, M. H., Hoare, C., Mangina, E., & O’Donnell, J. (2021). Review of urban building energy modeling (UBEM) approaches, methods and tools using qualitative and quantitative analysis. *Energy and Buildings*, *246*, 111073.

Ascione, F., Bianco, N., De Stasio, C., Mauro, G. M., & Vanoli, G. P. (2017). Artificial neural networks to predict energy performance and retrofit scenarios for any member of a building category: A novel approach. *Energy*, *118*, 999–1017.

Athey, S. (2017). Beyond prediction: Using big data for policy problems. *Science*, *355*(6324), 483–485.

Battocchi, K., Eleanor, D., Maggie, H., Greg, L., Paul, O., Miruna, O., & Vasilis, S. (2019). *EconML: A Python Package for ML-Based Heterogeneous Treatment Effects Estimation*. EconML: A Python Package for ML-Based Heterogeneous Treatment Effects Estimation.

Beagon, P., Boland, F., & Saffari, M. (2020). Closing the gap between simulation and measured energy use in home archetypes. *Energy and Buildings*, *224*, 110244.

Belaïd, F., & Massié, C. (2023). Driving forward a low-carbon built environment: The impact of energy context and environmental concerns on building renovation. *Energy Economics*, *124*, 106865.

Blöbaum, P., Götz, P., Budhathoki, K., Mastakouri, A. A., & Janzing, D. (2024). DoWhy-GCM: An extension of DoWhy for causal inference in graphical causal models. *Journal of Machine Learning Research*, *25*(147), 1–7.

Bontempi, G., & Flauder, M. (2015). From dependency to causality: A machine learning approach. *J. Mach. Learn. Res.*, *16*(1), 2437–2457.

Caliendo, M., & Kopeinig, S. (2008). Some practical guidance for the implementation of propensity score matching. *Journal of Economic Surveys*, *22*(1), 31–72.

Camarasa, C., Mata, É., Navarro, J. P. J., Reyna, J., Bezerra, P., Angelkorte, G. B., Feng, W., Filippidou, F., Forthuber, S., & Harris, C. (2022). A global comparison of building decarbonization scenarios by 2050 towards 1.5–2 C targets. *Nature Communications*, *13*(1), 3077.

Chen, S., Zhang, G., Xia, X., Chen, Y., Setunge, S., & Shi, L. (2021). The impacts of occupant behavior on building energy consumption: A review. *Sustainable Energy Technologies and Assessments*, *45*, 101212.

Chen, X., Abualdenien, J., Singh, M. M., Borrmann, A., & Geyer, P. (2022). Introducing causal inference in the energy-efficient building design process. *Energy and Buildings*, *277*, 112583.

Chernozhukov, V., Chetverikov, D., Demirer, M., Duflo, E., Hansen, C., Newey, W., & Robins, J. (2018). Double/debiased machine learning for treatment and structural parameters. *The Econometrics Journal,* Volume 21, Issue 1, 1 February 2018, Pages C1–C68.

Chuah, J. W., Raghunathan, A., & Jha, N. K. (2013). ROBESim: A retrofit-oriented building energy simulator based on EnergyPlus. *Energy and Buildings*, *66*, 88–103.

Cook, T. D., Campbell, D. T., & Shadish, W. (2002). *Experimental and quasi-experimental designs for generalized causal inference* (Vol. 1195). Houghton Mifflin Boston, MA.

Crawley, D. B., Lawrie, L. K., Winkelmann, F. C., Buhl, W. F., Huang, Y. J., Pedersen, C. O., Strand, R. K., Liesen, R. J., Fisher, D. E., & Witte, M. J. (2001). EnergyPlus: Creating a new-generation building energy simulation program. *Energy and Buildings*, *33*(4), 319–331.

Cunningham, S. (2021). *Causal inference: The mixtape*. Yale university press.

Curth, A., & Van der Schaar, M. (2021). Nonparametric estimation of heterogeneous treatment effects: From theory to learning algorithms. In *International Conference on Artificial Intelligence and Statistics* (pp. 1810-1818). PMLR.

D'Amico, B., Pomponi, F., Arehart, J. H., & Khaddour, L. (2025). Who cuts emissions, who turns up the heat? Causal machine learning estimates of energy efficiency interventions. *Energy and Buildings*, 116613.

Deb, C., Dai, Z., & Schlueter, A. (2021). A machine learning-based framework for cost-optimal building retrofit. *Applied Energy*, *294*, 116990.

Dermentzis, G., Ochs, F., Gustafsson, M., Calabrese, T., Siegele, D., Feist, W., Dipasquale, C., Fedrizzi, R., & Bales, C. (2019). A comprehensive evaluation of a monthly-based energy auditing tool through dynamic simulations, and monitoring in a renovation case study. *Energy and Buildings*, *183*, 713–726.

Eakins, J., Power, B., Ryan, G., Strömberg, H., & Diamond, L. (2025). Who saves energy and why? Analysing diverse behaviours in 27 European countries. *Energy Research & Social Science*, *121*, 103922.

Eurostat. (2025). *Energy consumption in households*. https://ec.europa.eu/eurostat/statistics-explained/index.php?title=Energy_consumption_in_households

Filippini, M., & Kumar, N. (2024). Determinants to the adoption of energy-efficient retrofits and the role of policy measures. *Applied Economics Letters*, *31*(10), 885–892.

Frisch, R., & Waugh, F. V. (1933). Partial time regressions as compared with individual trends. *Econometrica: Journal of the Econometric Society*, 387–401.

Garg, V., Mathur, J., & Bhatia, A. (2020). *Building energy simulation: A workbook using designbuilder™*. CRC Press.

Holland, P. W. (1986). Statistics and causal inference. *Journal of the American Statistical Association*, *81*(396), 945–960.

Imbens, G. W., & Rubin, D. B. (2015). *Causal inference in statistics, social, and biomedical sciences*. Cambridge university press.

International Energy Agency (IEA). (2023, July 12). *Tracking Clean Energy Progress 2023—Assessing critical energy technologies for global clean energy transitions*.

James, G., Witten, D., Hastie, T., & Tibshirani, R. (2013). *An introduction to statistical learning: With applications in R* (Vol. 103). Springer.

Jensen, P. A., Thuvander, L., Femenias, P., & Visscher, H. (2022). Sustainable building renovation–strategies and processes. *Construction Management and Economics*, *40*(3), 157–160.

Jiang, F., & Kazmi, H. (2025). What-if: A causal machine learning approach to control-oriented modelling for building thermal dynamics. *Applied Energy*, *377*, 124550.

Kaddour, J., Lynch, A., Liu, Q., Kusner, M. J., & Silva, R. (2022). Causal machine learning: A survey and open problems. *arXiv Preprint arXiv:2206.15475*.

Kirschbaum, J., Divkovic, D., & Meschede, H. (2025). From demand to action: Analysing building emissions and refurbishment scenarios towards climate neutrality. *Applied Energy*, *396*, 126302.

Künzel, S. R., Sekhon, J. S., Bickel, P. J., & Yu, B. (2019). Metalearners for estimating heterogeneous treatment effects using machine learning. *Proceedings of the National Academy of Sciences*, *116*(10), 4156–4165.

Kuzmanovic, M., Frauen, D., Hatt, T., & Feuerriegel, S. (2024). Causal machine learning for cost-effective allocation of development aid. In *Proceedings of the 30th ACM SIGKDD conference on knowledge discovery and data mining* (pp. 5283-5294).

Lee, M. (2016). *Matching, regression discontinuity, difference in differences, and beyond*. Oxford University Press.

Loga, T., Stein, B., Diefenbach, N., & Born, R. (2015). Deutsche Wohngebäudetypologie. *Beispielhafte Maßnahmen Zur Verbesserung Der Energieeffizienz von Typischen Wohngebäuden*, *2*.

Ma, Z., Cooper, P., Daly, D., & Ledo, L. (2012). Existing building retrofits: Methodology and state-of-the-art. *Energy and Buildings*, *55*, 889–902.

Molak, A. (2023). *Causal Inference and Discovery in Python: Unlock the secrets of modern causal machine learning with DoWhy, EconML, PyTorch and more*. Packt Publishing Ltd.

Moran, F., Blight, T., Natarajan, S., & Shea, A. (2014). The use of Passive House Planning Package to reduce energy use and CO2 emissions in historic dwellings. *Energy and Buildings*, *75*, 216–227.

National Renewable Energy Laboratory (NREL). (2025). *EnergyPlus*. https://energyplus.net

Nguyen, A. T., Ahn, Y., Park, S., Park, S., & Pham, D. H. (2024). Meta learning regression framework for energy consumption prediction in retrofitted buildings: A case study of South Korea. *Journal of Building Engineering*, *96*, 110403.

Ossokina, I. V., Kerperien, S., & Arentze, T. A. (2021). Does information encourage or discourage tenants to accept energy retrofitting of homes? *Energy Economics*, *103*, 105534.

Passive House Institute. (2015). *PHPP – The Energy Balance and Passive House Planning Tool*. https://passivehouse.com/04_phpp/04_phpp.htm

Pearl, J. (2009). *Causality*. Cambridge university press.

Pearl, J. (2014). *Probabilistic reasoning in intelligent systems: Networks of plausible inference*. Elsevier.

Pedregosa, F., Varoquaux, G., Gramfort, A., Michel, V., Thirion, B., Grisel, O., Blondel, M., Prettenhofer, P., Weiss, R., & Dubourg, V. (2011). Scikit-learn: Machine learning in Python. *The Journal of Machine Learning Research*, *12*, 2825–2830.

Peñasco, C., & Anadón, L. D. (2023). Assessing the effectiveness of energy efficiency measures in the residential sector gas consumption through dynamic treatment effects: Evidence from England and Wales. *Energy Economics*, *117*, 106435.

Recart, C., & Dossick, C. S. (2022). Hygrothermal behavior of post-retrofit housing: A review of the impacts of the energy efficiency upgrade strategies. *Energy and Buildings*, *262*, 112001.

Rubin, D. B. (1974). Estimating causal effects of treatments in randomized and nonrandomized studies. *Journal of Educational Psychology*, *66*(5), 688.

Rubin, D. B. (1980). Randomization analysis of experimental data: The Fisher randomization test comment. *Journal of the American Statistical Association*, *75*(371), 591–593.

Saffari, M., & Beagon, P. (2022). Home energy retrofit: Reviewing its depth, scale of delivery, and sustainability. *Energy and Buildings*, *269*, 112253.
Schleich, J., Faure, C., & Meissner, T. (2021). Adoption of retrofit measures among homeowners in EU countries: The effects of access to capital and debt aversion. *Energy Policy*, *149*, 112025.
Seabold, S., & Perktold, J. (2010). Statsmodels: Econometric and statistical modeling with python. *SciPy*, *7*(1), 92–96.
Sharma, A., & Kiciman, E. (2020). DoWhy: An end-to-end library for causal inference. *arXiv Preprint arXiv:2011.04216*.
Tafti, A., & Shmueli, G. (2020). Beyond overall treatment effects: Leveraging covariates in randomized experiments guided by causal structure. *Information Systems Research*, *31*(4), 1183–1199.
Tafti, A., & Shmueli, G. (2025). Causal Inference Grounded in Causal Diagrams: Benefits, Limitations, and Opportunities for the Information Systems Field. *MIS Quarterly*, *49*(2), 409–428.
Thrampoulidis, E., Mavromatidis, G., Lucchi, A., & Orehounig, K. (2021). A machine learning-based surrogate model to approximate optimal building retrofit solutions. *Applied Energy*, *281*, 116024.
Umit, R., Poortinga, W., Jokinen, P., & Pohjolainen, P. (2019). The role of income in energy efficiency and curtailment behaviours: Findings from 22 European countries. *Energy Research & Social Science*, *53*, 206–214.
U.S. Energy Information Administration (EIA). (2024). *What is the consumption of energy in buildings in the United States?* https://www.eia.gov/tools/faqs/faq.php?id=86&t=1
von Zahn, M., Bauer, K., Mihale-Wilson, C., Jagow, J., Speicher, M., & Hinz, O. (2025). Smart green nudging: Reducing product returns through digital footprints and causal machine learning. *Marketing Science*, *44*(4), 954–969.
von Zahn, M., Güler, A., Pfeiffer, J., Reijers, H. A., & Hinz, O. (2026). Causal Machine Learning in Information Systems Research. *Business & Information Systems Engineering*. https://doi.org/10.1007/s12599-026-00999-x
Walsh, M. J. (1989). Energy tax credits and housing improvement. *Energy Economics*, *11*(4), 275–284.
Weber, T., Schneeberger, S., & Schuetz, P. (2025). Estimating Heterogeneous Treatment Effects of Building Energy Efficiency Retrofits Using Machine Learning. *Energy and Buildings*, 116369.
Zapata Gonzalez, D., Meyer, M., & Mueller, O. (2025). Bridging the gap between data-driven and theory-driven modelling–leveraging causal machine learning for integrative modelling of dynamical systems. *European Conference on Information Systems*.

# Appendix

Main setup parameters:

- **Location**: Binary climate indicator sampled from Binomial(1, 0.5), where 0 = moderate climate, 1 = cold climate
- **Household size**: Poisson($\lambda = 3$) distribution, representing number of occupants
- **Income**: Normal($\mu$ = \$50,000, $\sigma$ = \$15,000), truncated at zero to prevent negative values
- **Construction year**: Uniformly sampled from TABULA (Loga et al., 2015) building class periods spanning 1850-1994, with eight distinct classes (EFH A through EFH H)
- **Air density * specific thermal air capacity**: 0.34
- **Full load hours**: 1130
- **Norm temperatures**: Outside: -12.0 for cold and -5.0 for moderate climates; Inside: 20